\documentclass[namedreferences]{kluwer}

\usepackage{graphicx}
\begin{document}
\begin{article}

\begin{opening}
\title{The impact of chemical evolution on the observable properties of
stellar populations}

\author{Monica \surname{Tosi}\email{tosi@bo.astro.it}}
\institute{Osservatorio Astronomico, Bologna, Italy}  

\runningtitle{Chemical evolution and stellar populations}
\runningauthor{Tosi}

\begin{abstract} 
The major effects of the chemical evolution of galaxies on the characteristics
of their stellar populations are reviewed. A few examples of how the
observed stellar properties derived from colour--magnitude diagrams
can constrain chemical evolution models are given.
\end{abstract}



\end{opening}

\section{Introduction}
The chemical evolution of galaxies has a strong impact on
the observable properties of their stellar populations.
In this review, I briefly summarize the main effects 
on the major stellar features,
and give a few examples of observable properties which can, in turn,
constrain the chemical evolution parameters.
In particular, I will try to show how the colour--magnitude diagrams (CMDs) of
observed stellar populations can provide useful information to derive their
initial mass function (IMF), to infer the evolution of the abundance 
gradients in the Galaxy, and to derive the star formation (SF) histories 
of dwarf galaxies.

\begin{figure}
\centerline{\includegraphics[width=17pc]{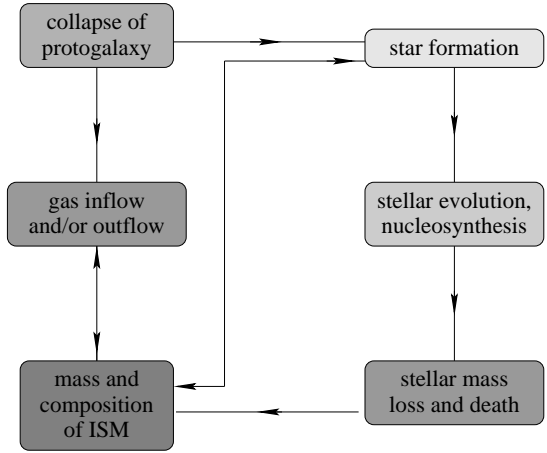}}
\caption{General scheme for the chemical evolution of a galaxy.}
\label{scheme}
\end{figure}

The general scheme for the chemical evolution of a galaxy was drawn
by Beatrice Tinsley (1980) already 20 years ago and is shown in
Fig.\ref{scheme}.
When a galaxy forms, independently 
of the formation scenario, after a while it starts to form stars. These
stars evolve and synthesize in their interiors heavier and heavier elements,
which are then ejected in the surrounding medium when the stars lose mass
and die. In this way they pollute the interstellar medium (ISM) and modify
both its mass and chemical composition. 
Meanwhile, the ISM mass and metallicity can be affected also by other
phenomena, such as gas exchanges with adjacent regions (gas losses
or accretions, or both). Hence, the next generation of stars formed
in the region have a somewhat different initial composition from the
previous generation and, therefore, a slightly different evolution.
Depending on the morphological type of the considered galaxy, many cycles 
of this kind can occur during its lifetime.  Models for galaxy
chemical evolution are the tool to take all of them into account,
as well as their effects on the stellar and gaseous properties.

Given our poor knowledge of the actual 
physical mechanisms regulating the above phenomena, what is usually done to
compute chemical evolution models is to parametrize the phenomena with fairly
simplistic laws. The major parameters are the SF law, the IMF, 
the gas flows in and out of the considered region, and all
the quantities involved in stellar nucleosynthesis (e.g. stellar lifetimes,
mass loss, opacities, treatment of convection, etc.).

All these parameters have important implications on the stellar properties. 
In particular:
\begin{itemize}[$\bullet$]
\item The SF affects the age (in the sense that the earlier the SF onset, the 
  older the stellar population), the number (the higher the SF rate, the higher
  the number of formed stars), the colours (the more recent the SF activity,
  the bluer the stellar colours), and the element abundances (the higher the
  SF activity, the higher the heavy element production).
\item The IMF affects the relative numbers of stars of different mass (flatter
  IMFs imply higher ratios of high/low mass stars than steeper ones), and
  the abundance ratios of elements produced by stars in different mass
  ranges (e.g. the N/O ratio, since nitrogen is mostly produced by intermediate
  mass stars and oxygen is produced only by massive stars).
\item The stellar nucleosynthesis is directly related to the stellar chemical
  composition.
\item Gas flows have more indirect, but not less important, effects: they can
  modify the chemical abundances of the ISM and hence the initial metallicity 
  of the stars formed from it; and they influence the SF (strong gas
  losses, like galactic winds, may remove all the gas from a region thus
  inhibiting any further SF activity there, or, viceversa, gas accretion
  may allow to reach the gas density necessary to induce SF).
\end{itemize}

In this review, I focus on stellar observable properties affected
by the SF and the IMF, and derivable from the CMDs.

\begin{figure}
\centerline{\includegraphics[width=18pc]{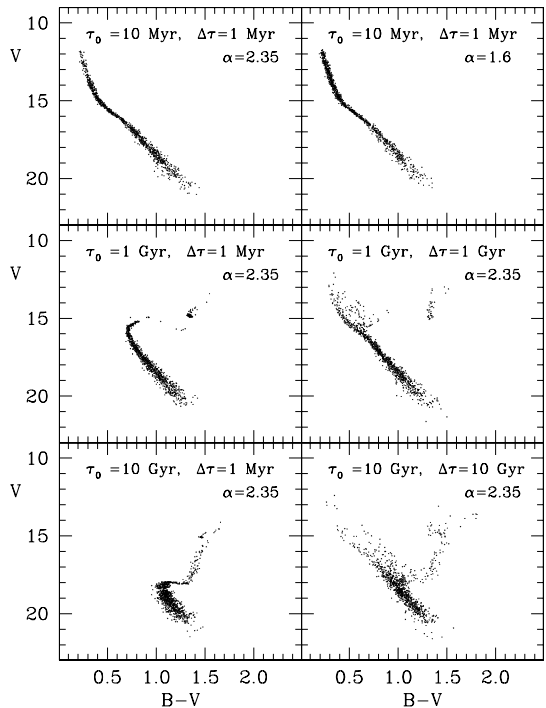}}
\caption{Synthetic CMDs for various SF histories and IMF (see text for
details), based on the Padova stellar evolution models with Z=0.02
(Bressan et al. 1993). Indicated in each panel are the IMF slope $\alpha$,
the starting epoch $\tau$ (backwards from present) and the duration 
$\Delta\tau$ of the SF activity. }
\label{granadasim}
\end{figure}

How do the SF and IMF influence the morphology of the CMD of a stellar
population~? For an immediate understanding, let's 
assume to have a hypothetical stellar system, with
known distance and reddening [e.g. (m-M)$_0$=12.5 and E(B-V)=0.45, 
appropriate for a Galactic open cluster], and to observe it with a
telescope allowing us to resolve 1000 of its stars, with a fairly small
photometric error and a derivable incompleteness. The various panels of
Fig.\ref{granadasim} show the expected appearance of the CMD of the 1000
resolved stars for different SF histories and different IMF slopes.

If the SF has started 10 Myr ago and
lasted 1 Myr, with a Salpeter's IMF ($\alpha$=2.35), the CMD will be as in
the top left panel, with all the visible stars still on the main sequence (MS).
If the SF has been the same, but the IMF was much flatter than Salpeter's 
(e.g. $\alpha$=1.6), the CMD will be as in the top right panel, with a 
stellar distribution 
definitely skewed toward more massive, more luminous objects (an effect
which will be more evident in the luminosity function, LF).
If the SF activity has started 1 Gyr ago, depending on its
duration it can lead to the CMD of the middle left panel (for a SF
duration of 1 Myr) or to that of the middle right panel (for a duration
of 1 Gyr and a constant rate until now). In both cases we can recognize
the MS turnoff of the 1 Gyr old stars, the subgiant and red giant branches
and the clump of core-He-burning objects. The right hand panel in addition
contains younger
stars on the MS and at the blue edge of the core-He-burning loops.
Finally, if the onset of the SF has occurred as early as 10 Gyr ago, the
CMD will be as shown in the two bottom panels: the left one in the case of
a 1 Myr burst and the right one in the case of a constant rate till the present
epoch. Again, in the instantaneous case we essentially see only the
10 Gyr isochrone, while the continuous SF allows also for the presence of
younger, more massive stars, concentrated on the evolutionary phases of 
relatively longer duration: MS, red giant and asymptotic branches, 
and hot and cool edges of the blue loops.

Multiple episodes of SF activity would lead to CMDs which are essentially
the combination of these simple ones. It is thus evident how the
SF history and the IMF can be derived from an appropriate interpretation of 
empirical CMDs. This is the reason why the diagrams are so widely used to
infer the evolutionary properties of stellar populations. In the following,
three applications of the CMDs to constrain chemical evolution
models are described.

\section{Derivation of the IMF}

\begin{figure}
\centerline{\includegraphics[width=3.0in]{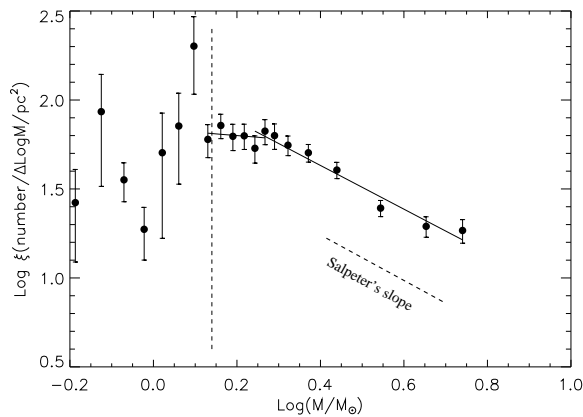}}
\caption{IMF in the cluster R136 of the LMC (Sirianni et al. 2000). }
\label{imf} 
\end{figure}

Thanks to modern instrumental improvements, it is now
possible to derive from CMDs and LFs the IMF of star clusters down to
very low masses. HST has significantly contributed to this goal, both with
optical and infrared imaging, allowing people to resolve single objects
even at quite faint magnitudes. It has thus been possible to derive the
IMF in globular clusters (see e.g. Piotto et al. 1997, De Marchi et al. 
2000 for NGC 6397) and open clusters (see e.g. Luhman 2000 for Trapezium,
$\rho$ Oph and IC 348) 
of our Galaxy down to masses as low as 0.07 M$_{\odot}$. The inferred functions
are fairly similar to each other and show a flat slope up to 
$\sim$0.6 M/M$_{\odot}$ and a power law slope similar to, or slightly steeper
than, Salpeter's for more massive stars.
The resolving power of HST has allowed Sirianni et al. (2000) to infer the 
IMF even in the LMC. In the cluster R136 they have reached masses down 
to 0.6 M$_{\odot}$ (Fig.\ref{imf}), finding a
roughly Salpeter's slope for stars more massive than $\sim$1.8 M$_{\odot}$
and a flattening towards lower masses.

When a significantly large sample of IMFs derived from clusters of different
ages, metallicities and environments will be available, we will better
understand if the IMF is indeed roughly universal, as currently suggested
by several studies, and what is its actual slope in all the mass ranges:
an achievement of great importance in the field of chemical evolution
modeling.

\section{Open clusters as tracers of the evolution of the abundance gradients}

One of the important questions that still remain without satisfactory answer
concerns the evolution of abundance gradients in our and in other galaxies.
The distribution of heavy elements with galactocentric distance, as
derived from young objects like HII regions or B stars, shows a steep 
negative gradient in the disk of the Galaxy and of other well studied spirals. 
Does this slope change with time or not ? And, in case,
does it flatten or steepen ? 

Unfortunately, Galactic chemical evolution models do not provide a consistent
answer: even those that are able to reproduce equally well the largest set
of observational constraints predict different scenarios for the early epochs.
By comparing with each other the most representative models,
Tosi (1996) showed that the predictions on the gradient evolution range
from a slope initially positive which then becomes negative and steepens with
time, to a slope initially negative which remains roughly constant, to a 
slope initially negative and steep which gradually flattens with time (see
also Tosi 2000 for updated discussion and references).

From the observational point of view, the situation is not much clearer.
Planetary Nebulae (PNe) show a gradient similar to that derived from
HII regions, possibly slightly flatter than the latter, when the sample 
is restricted to PNe of type II, which are supposed to have progenitors 
formed on average 2-3 Gyr ago. Data on field stars are inconclusive, due
to the large uncertainties affecting the older, metal poorer ones. 
\begin{figure}
\centerline{\includegraphics[width=14pc]{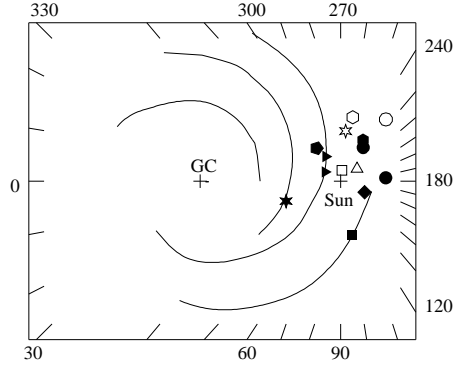}}
\caption{Location in the Galactic plane of the open clusters already
observed for the project on the gradient evolution. The Sun and the Galactic
center positions are marked with crosses. The solid curves represent 
the main spiral arms.
}
\label{open}
\end{figure}
Open clusters probably represent the best tool to understand whether and how
the gradient slope changes with time, since their ages, distances and 
metallicities are more safely derivable than for field stars, specially 
at large distances from us. However, also the data on open clusters available 
so far are inconclusive, as shown by Bragaglia et al. (2000) using the
compilation of ages, distances and metallicities listed by Friel (1995).
By dividing her clusters in four age bins, we find in fact no significant
variation in the gradient slope, but we do not know if this reflects
the actual situation or the inhomogeneities in the data treatment of 
clusters taken from different literature sources.

For this reason, we are undertaking a long term project of accurate
photometry and high-resolution spectroscopy to derive homogeneously ages, 
distances, reddening and element abundances in open clusters of various
ages and galactic locations and eventually infer from them the  gradient
evolution (see Sandrelli et al. 1999, and references therein). 
Up to now we have acquired the photometry of 15 clusters and
the spectroscopy of 3 of them. Fig.\ref{open} shows the Galactic
location of the clusters for which we have already derived the CMDs.

\section{Star formation histories in dwarf galaxies}

Another hot topic in the field of galaxy evolution is the
SF regime in dwarf galaxies of late morphological type: is it bursting
(i.e. taking place in short and intense episodes of activity separated by long
quiescent phases), gasping (i.e. in long episodes of moderate
activity separated by short quiescent intervals), or continuous ?

Searle et al. (1973) suggested that the observed features of blue compact
galaxies could be explained only if they have a bursting regime of SF,
a suggestion confirmed by the chemical evolution models computed by e.g.
Matteucci \& Tosi (1985) and Pilyugin (1993), who found
that these galaxies should experience no more than 7-10 bursts during their
life. Other authors (e.g. Carigi et al. 1995, Larsen et al. 2000), however, have 
suggested that also a continuous low SF activity could reproduce
equally well the observed chemical abundances, even in the most extreme
case of IZw18 (Legrand 2000), the most metal poor galaxy discovered so far.

This inconsistency between chemical evolution models of different groups
is due to the fact that the models are poorly constrained for most of 
the external galaxies, where only data on the gas and total masses and
on the chemical abundances derived from HII regions are available.

The only direct way to derive the SF regime is to study the resolved stellar 
populations and infer the epochs and intensities of the SF activity from 
their CMDs. To this aim, we (Tosi et al. 1989, 1991, Greggio et al. 1998) 
developed the method of synthetic CMDs which has also been adopted and 
modified by other groups (e.g. Tolstoy \& Saha 1996, Aparicio et al. 1996)
and is now widely used. Eva Grebel (this volume) has reviewed the SF 
histories derived 
for Local Group dwarfs, but now that powerful telescopes like HST and the
ESO-VLT are available, the method can be applied successfully also to more
distant galaxies, as described by several authors at this Conference.

\begin{figure} [H]
\centerline{%
\begin{tabular}{c@{\hspace{0.01pc}}c}
\includegraphics[width=2.25in]{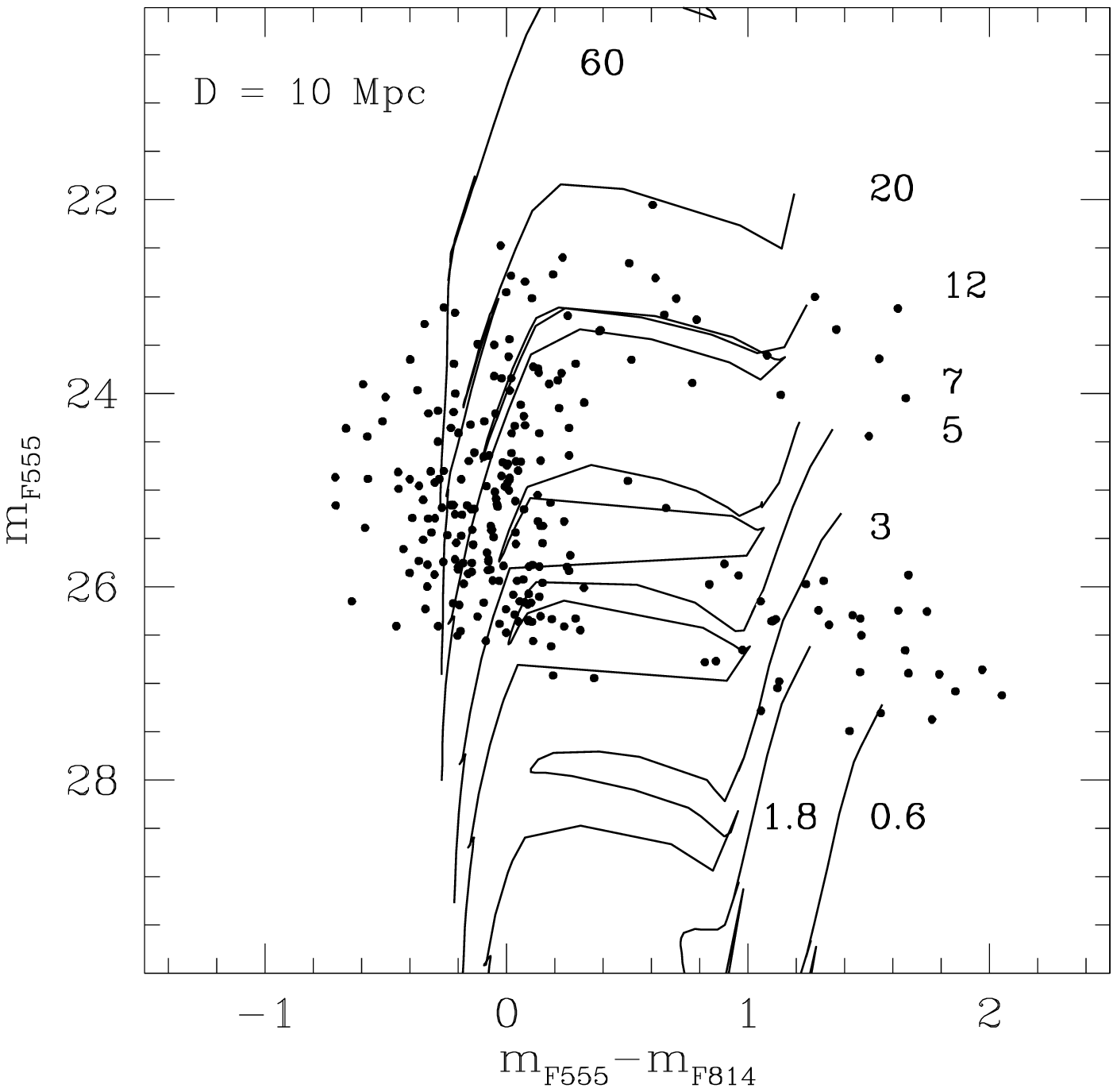} &
\includegraphics[width=2.35in]{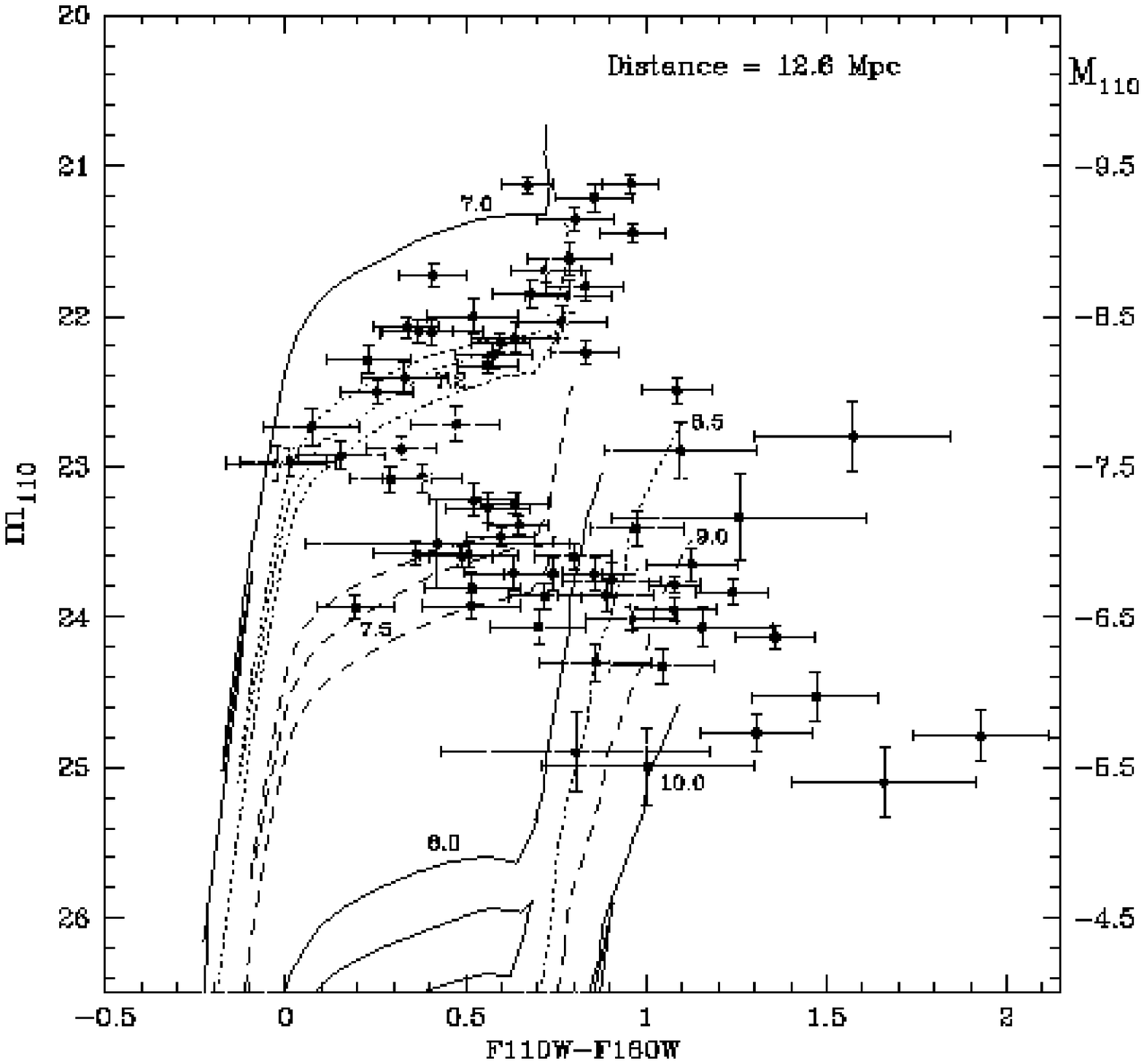} 
\end{tabular}}
\caption{Optical and NIR CMD of the resolved stars in the main body of
IZw18, as resulting from HST data (Aloisi et al. 1999 and Ostlin 2000,
respectively). The masses indicated in the left panel are in M$_{\odot}$.}
\label{izw18cmd}
\end{figure}

The important implications that CMDs can have on chemical evolution models
of fairly distant galaxies can be examplified by discussing the 
interesting case of IZw18.
Until very recently, this blue compact galaxy was considered the
prototype of a really young system, experiencing now its first burst of
star formation. This youth was considered the best (possibly the only)
way to explain its extremely low metallicity (Z=1/50 Z$_{\odot}$). A few years
ago, Hunter \& Thronson (1995) and Dufour et al. (1996) obtained 
HST-WFPC2 images of IZw18, allowing for the first time to
resolve a few hundreds of its individual stars. More recently Ostlin (2000)
has obtained HST-NICMOS images. The optical and infrared data sets have
provided the CMDs shown in Fig.\ref{izw18cmd}.
Notice that on the optical CMD we
have overimposed the Padova stellar evolutionary tracks of the appropriate
metallicity (Z=0.0004), adopting a distance of 10 Mpc, while
on the infrared CMD Ostlin has overimposed the Geneva isochrones of
the same metallicity, adopting a distance of 12.6 Mpc. 

Both diagrams show two groups of stars: a brighter, bluer group, corresponding
to stars of $\gsim$ 12 M$_{\odot}$, born during a recent burst occurred around 
12-15 Myr
ago (the exact values depending of course on the adopted distance), and a
faint, red group, corresponding to stars with mass lower than 3 M$_{\odot}$
and age between 0.3 and 15 Gyr. Hence, an old stellar population definitely
exists in IZw18 and rules out the hypothesis of one single and recent
SF burst (see Aloisi et al. 1999 and Ostlin 2000 for details). 

\begin{figure}
\centerline{\includegraphics[width=28pc]{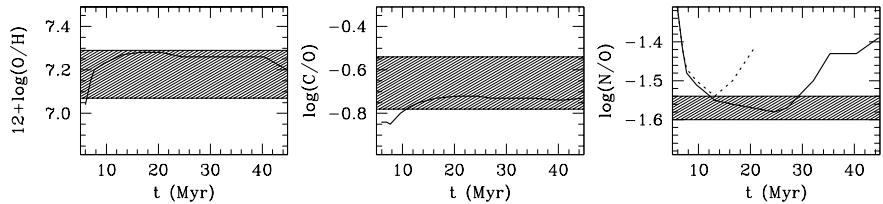}}
\caption{Comparison of the predictions of the chemodynamical evolution
models for IZw18 by Recchi with the C, N and O abundances derived from
HII region data.}
\label{recchi}
\end{figure}

This result puts strong constraints on chemical evolution models
and it is now important to check if we can reproduce the observed masses
and chemical abundances of IZw18, with the SF and IMF derived 
from the empirical CMDs. Aloisi et al. (this volume) present the 
preliminary results of
one of such attempts. Since Recchi et al. (2000) have recently
proposed a chemo-dynamical model for IZw18, taking into account the
effects on the ISM of the explosions of Supernovae of types Ia and II, 
I asked them to run their code with the SF history and rates derived 
by Aloisi et al.'s (1999), 
i.e. with two episodes of SF, the first one occurring 300 Myr before 
the second one and with a SF rate 10 times lower. Fig.\ref{recchi}
shows the predictions of such models for oxygen, C/O and N/O as a function
of the time elapsed from the onset of the second burst. The shaded regions
correspond to the ranges of abundances derived by several authors (see
references in Recchi et al. 2000). The two curves in the N/O panel correspond
to assuming (dotted line) or not (solid line) primary nitrogen production 
in massive stars. It is apparent that all the solid curves fit quite well the
data, in the time interval between 12 and 28 Myr, thus supporting both
the compatibility of two bursts of SF with the low observed abundances and
the burst properties (age, intensity, etc.) derived from the CMDs.

I believe that from now on, the most correct way to get a better
insight on the SF regime and on the whole evolutionary scenario of late
type dwarfs is to infer the SF history and IMF from the CMDs of 
their resolved stellar populations and use this information as
input data for chemical evolution models, to check the consistency with 
their observed masses and chemistry.

\begin{acknowledgements}
I thank Simone Recchi for computing new IZw18 models. 
This work has been partially supported by the Italian MURST, through
COFIN98 at Arcetri and by the Organizers of this enjoyable meeting.

\end{acknowledgements}

\end{article}
\end{document}